\documentclass{entcs} 

\usepackage{bonite}

\def\lastname{Hirschkoff, Hirschowitz, Hym, Pardon, Pous}

\begin{document}

\begin{frontmatter}

  \title{Encapsulation and Dynamic Modularity
    \\
    in the $\pi$-calculus}
\thanks{This work has been supported by the French project ANR Arassia
    \textit{``Modularit{\'e} Dynamique Fiable''}.}
  
  \author{Daniel Hirschkoff, Aur{\'e}lien Pardon}%
  \address{ENS Lyon, LIP, Universit{\'e} de Lyon, CNRS, INRIA}
  \author{Tom Hirschowitz}%
  \address{LAMA, Universit{\'e} de Savoie, CNRS}
  \author{Samuel Hym}%
  \address{LIFL, Universit{\'e} de Lille 1, INRIA, CNRS}
  \author{Damien Pous}%
  \address{SARDES, LIG, Grenoble, CNRS, INRIA}

  \begin{abstract}
    We describe a process calculus featuring
    high level constructs for component-oriented programming in a
    distributed setting.
    We propose an extension of the higher-order $\pi$-calculus
    intended to capture several important mechanisms related to
    component-based programming, such as dynamic update,
    reconfiguration and code migration.
    In this paper, we are primarily concerned with the possibility to
    build a distributed implementation of our calculus. Accordingly,
    we define a low-level calculus, that describes how the high-level
    constructs are implemented, as well as details of the data
    structures manipulated at runtime.
%
    We also discuss current and future directions of research in
    relation to our analysis of component-based programming.
  \end{abstract}

  \begin{keyword}
    Distributed computing, component-based programming, process
    algebra, higher-order calculi, abstract machine.
  \end{keyword}
  
\end{frontmatter}

\section{A Core Calculus for Dynamic Modularity}
\label{sec:design}

\subsection{Motivations of this Work}

This paper describes work on component-oriented programming and the
$\pi$-calculus. Our long term goal is the design and implementation of
a prototype programming language meeting the following requirements.
\begin{itemize}
\item\label{req-pi} It should be suitable for \emph{concurrent,
    distributed programming}. For instance, usual distributed,
  parallel algorithms should be easily implementable, as well as
  lower-level communication infrastructure for networks. Furthermore,
  it should enjoy a well-understood and tractable behavioural theory.
\item\label{req-mod} It should provide constructs for
  \emph{modularity}, in the standard, informal sense that programs
  should be built as an assembly of relatively independent computation
  units (or \emph{modules}) interacting at explicit interfaces.
  Moreover, modularity should come with \emph{encapsulation} features,
  e.g., it should be possible to exchange two modules implementing the
  same interface without affecting the rest of the code.
\item\label{req-mody} The modular structure of programs should be
  available at execution time, so as to ease standard dynamic
  operations such as migration, dynamic update, or passivation of
  modules.  We call this requirement \emph{dynamic modularity}. 
  The notion of dynamic modularity gathers the most challenging
  features of component based programming we are interested in
  modelling and analysing.
\item\label{req-comp} Finally, we are seeking a reasonably
  implementable language, at least permitting rapid prototyping of
  distributed applications. 
\end{itemize}



In this paper, we describe our proposal for a core process algebra to
represent and analyse dynamic modularity (this section). In designing
this formal model, we have put a stress on the possibility to deploy
and execute processes in a distributed fashion. In
Sect.~\ref{sec:implem}, we describe our prototype implementation by
defining an abstract machine for the distributed execution of
processes. This description is given via the definition of a new
process calculus, where low-level aspects related to the
implementation (notably, communication protocols), as well as the data
structures at work in the machine, are made explicit.

\subsection{Syntax and Semantics of \kepi}

In order to provide a formal treatment of the questions described
above, we study an extension of the higher-order $\pi$-calculus,
called \kepi. 
We choose the $\pi$-calculus for two main reasons: first,
message-based concurrency seems an appropriate choice to define a
model for concurrent programming at a reasonable level of abstraction.
Second, working in the setting of a process algebra like the
$\pi$-calculus makes it possible to define a core formalism in which
we can analyse the main questions, both theoretical and pragmatic,
related to the implementation of primitives for dynamic modularity.
Third, we might hope for this to benefit from the considerable amount
of research that has been made on $\pi$-calculus based formalisms.

Our calculus inherits ideas from numerous previous studies, among
which~\cite{HildebrandtGM04,hennessy:book:dpi,phd:fournet}, and in
particular the Kell calculus~\cite{BidingerS03,Schmitt05}.
The grammar for processes is given in Fig.~\ref{fig:kp:syntax} 
-- we suppose two infinite sets of names ($a,b,m,n,k,x$)
and process variables $(X,Y)$. 
In addition to the usual $\pi$-calculus constructs, we have
\emph{modules}, $n[P]$, which can be seen as located processes (note
that modules can be nested). $M.P$ is a process willing to emit
(first- or higher-order) message $M$ and then proceed as $P$. $R\tri
P$ stands for a process willing to acquire a resource: this can mean
either receiving a first- or higher-order message (cases $a(x)$ and
$a(X)$, respectively), or passivating a module.
The input prefixes and restriction are binding constructs, and we
write \fn{P} for the free names of process $P$. $\{b/x\}$ (resp.\
$\{P/X\}$) denotes the capture avoiding substitution of name $x$
(resp.\ process variable $X$) with name $b$ (resp.\ process $P$).

Notice that a process variable $(X)$ does not form a process by
itself: it has to be enclosed in a module $(n[X])$ or a message
$(\outm a X.P)$. This restriction mainly ensures that the content of
several modules cannot be merged into a single module; this helps
simplifying the implementation of the calculus.  
%
\begin{figure}[t]
  \centering
  \begin{align*}
    P,Q &::=~ P\,|\,P \OR (\new n)P \OR \nil\OR n[P]\OR n[X]\OR M.P \OR R\tri P 
    \tag{\textrm{processes}}\\
    M &::=~ \outm{a}{n} \OR \outm{a}{P} \OR \outm{a}{X}
    \tag{\textrm{output prefixes}}\\
    R &::=~ a(x)\OR a(X)\OR n[X] 
    \tag{\textrm{input prefixes}}\\
    \E &::=~ []\OR \E\,|\,P \OR (\new n)\E \OR n[\E]
    \tag{\textrm{evaluation contexts}}
  \end{align*}
  \caption{\kepi: Syntax for processes, evaluation contexts}
\label{fig:kp:syntax}
\end{figure}

\begin{figure}[t]
  \centering
    \begin{mathpar}
      \inferrule*[left=\RuleKComm]{a,b\notin\capt{\E_1}\cup\capt{\E_2}}
      {\E[\E_1[\outm a b.P]~|~\E_2[a(x)\tri Q]] \red
        \E[\E_1[P]~|~\E_2[Q\{b/x\}]]}
      \\
      \inferrule*[left=\RuleKHOComm]{\fn{P'}\cap(\capt{\E_1}\cup\capt{\E_2})=\emptyset}
      {\E[\E_1[\outm a {P'}.P]~|~\E_2[a(X)\tri Q]] \red
        \E[\E_1[P]~|~\E_2[Q\{P'/X\}]]}
      \\
      \inferrule*[left=\RuleKPass]{ }{ \E[n[P]~|~n[X]\tri Q] \red \E[Q\{P/X\}]}
      \\
      \inferrule*[left=\RuleKCongr]{ P'\equiv P\and P\red Q\and Q\equiv Q'}{ P'\red Q'}
    \end{mathpar}
  \caption{Operational Semantics of \kepi}
\label{fig:kp}
\end{figure}


Fig.~\ref{fig:kp} presents the rules defining the reduction relation,
$\red$. It makes use of evaluation contexts, which are processes with
a hole that does not occur under a prefix (Fig.~\ref{fig:kp:syntax}).
Given an evaluation context $\E$, we define the set of \emph{captured
  names} in $\E$, written $\capt{\E}$, as the set of names that are
bound at the occurrence of the hole in $\E$. The definition of $\red$
makes use of structural congruence, $\equiv$, the least congruence
satisfying the following axioms:
\begin{mathpar}
  P|\nil\equiv P
  \and
  P|Q\equiv Q|P
  \and
  P|(Q|R)\equiv (P|Q)|R
  \\
  (\new c)(\new d)P\equiv (\new d)(\new c)P
  \and
  (\new c)(P|Q)\equiv P\,|\,(\new c)Q\text{ if }c\notin\fn{P}
  \and
  (\new c)\nil\equiv\nil
\end{mathpar}

\begin{figure}
  \centering
  \includegraphics[width=.5\linewidth]{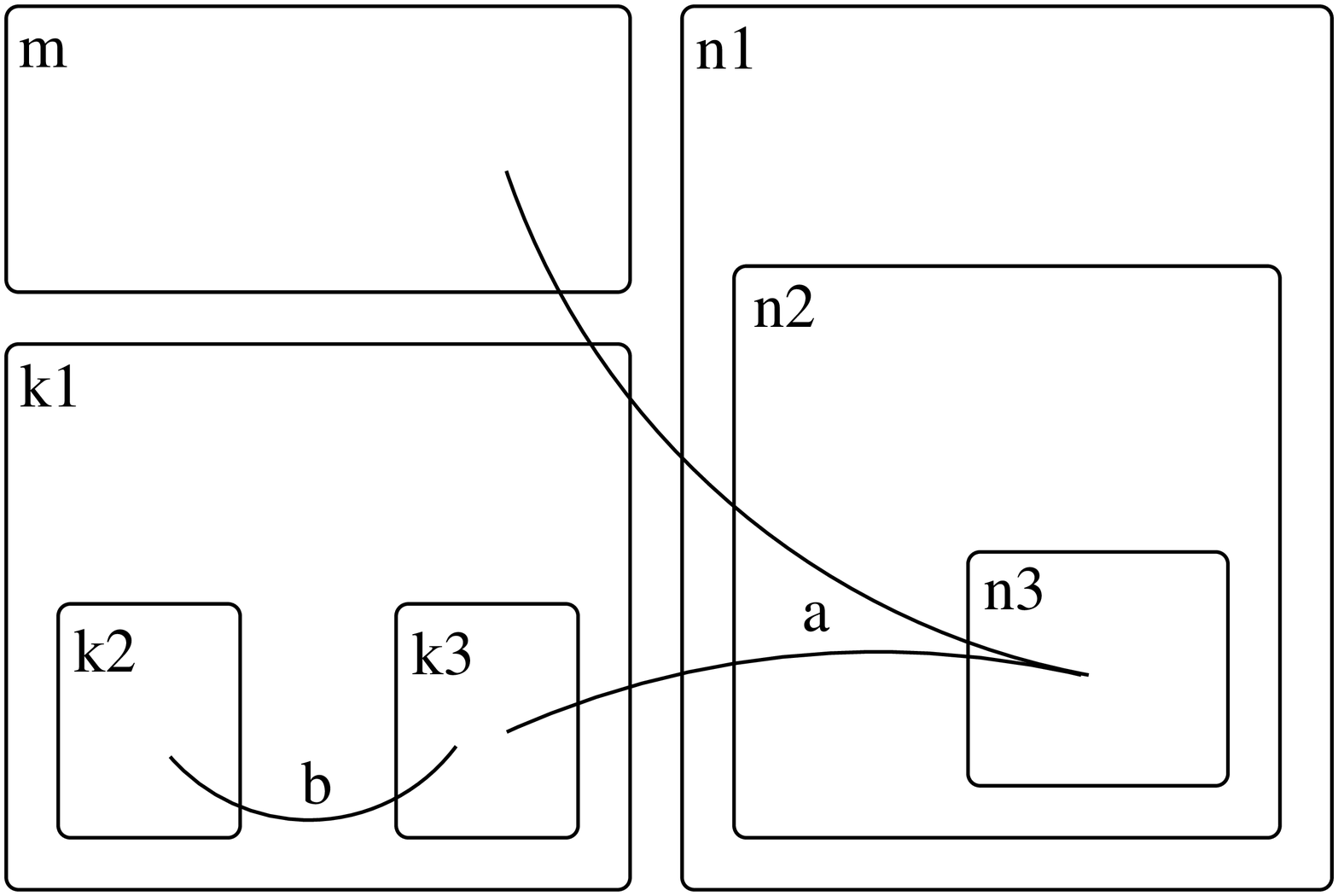}
  \caption{A configuration.}
  \label{fig:uml}
\end{figure}

\medskip


\paragraph{An example \kepi{} process.}

A \kepi{} term represents a configuration consisting of a hierarchy of
modules, in which processes are executed.  To illustrate the
mechanisms at work in \kepi, we briefly discuss an example
configuration. Given some \kepi{} processes $P,P'_i,Q_i$
Fig.~\ref{fig:uml} depicts a process of the form
\begin{align*}
  (\new a)(m[P]~|~n_1[P'_1~|~n_2[P'_2~|~n_3[P'_3]]]
  ~|~ k_1[(\new b)(Q_1~|~k_2[Q_2]~|~k_3[Q_3])]) 
\end{align*}
 (note that the localisation of
the restrictions on $a$ and $b$ does not appear on the picture).
There are basically two forms of interaction in \kepi:
\emph{communication} and \emph{passivation}.
Communication involves the transmission of a name or a process; 
it is \emph{distant}, in the sense that a process $\outm{a}{b}.P$ can
synchronise with a receiver $a(x)\tri Q$ sitting in a different
location, provided they share the name $a$. On the picture, a process
running in $k_3$ can exchange messages with another one in $n_3$
(using channel $a$), as well as with a third process running in $k_2$
(using channel $b$).
On the contrary, passivation is local: only a process running at $n_1$
is able to passivate module $n_2$. This is described by the third
axiom of Fig.~\ref{fig:kp}: up to structural congruence, the module
being passivated and the process that takes control over it must be in
parallel.
Passivation is a central construct in our formalism, and can be used
to implement very different kinds of manipulations related to dynamic
modularity.  For instance, taking $Q=n'[X]$ in the above reduction
leads to a simple operation of module renaming; with $Q=\outm{c}{X}$,
the module $n$ will be `frozen' and marshalled into a message to be
sent on channel $c$; finally, taking $Q=n[X]\,|\,n'[X]$ makes it
possible to duplicate a computation. When considering the actual
implementation of the behaviour of \kepi{} processes, it appears that
the last two examples clearly involve much more costly operations than
the first one.

\subsection{On the Design Choices in the Definition of \kepi}

\subsubsection{Restricted names as localised resources}

An important commitment that we make in the design of \kepi{} is that,
contrarily to several existing proposals, we do \emph{not} allow
channel names to be extruded across module boundaries: neither do we
include an axiom of the form $n[(\new b)\,P]\,\equiv\,(\new b)\,n[P]$
in structural congruence, nor do we implement name extrusion across
modules along reduction steps that would require it.
%
%
%
%

This design choice is related to the notion of module that we put
forward in \kepi: if we were to allow name extrusion, some processes
would admit two legitimate but very different reduction paths.  Take
indeed
\begin{mathpar}
  P\quad \eqdef\quad m[(\new a)\outm b a.P]~|~b(x).Q~|~m[X]\tri
  (m_1[X]~|~m_2[X])
  \enspace.
\end{mathpar}
We would have the following sequences of reductions:
\begin{align*} 
  P &\to (\new a)(m[P]~|~Q\{a/x\}~|~m[X]\tri (m_1[X]~|~m_2[X]))
  \\
  &
  \to (\new a)(Q\{a/x\}~|~m_1[P]~|~m_2[P])  \\
  & \text{and}\\
  P &\to b(x).Q ~|~ m_1[(\new a)\outm b a.P]~|~m_2[(\new a)\outm b
  a.P]
  \\
  & \to (\new a)(Q\{a/x\}~|~m_1[P])~|~m_2[(\new a)\outm b a.P]
\end{align*}
We claim that none of these paths is fully satisfactory, and that such
a situation should be avoided -- the choice between these two
behaviours being left to the user, through explicit programming.
Therefore, we interpret the names declared inside a module as private
resources, that should remain local to that module.
%
%
Passivating module $n$ hence means getting hold of the local
computations, as well as of the resources allocated in $n$. Typically,
names allocated in module $n$ can be viewed either as temporary
resources allocated for the computations taking place at $n$, or as
methods provided for sub-modules of $n$, for which $n$ acts as a
library.

As a consequence, the user is made aware of the localisation of
resources; this choice also helps considerably in the implementation
of \kepi, essentially because we always know how to route messages to
channels (see below; a similar idea is present in existing
implementations of $\pi$-calculus-related process algebras, such
as~\cite{JoCaml}).
At the same time, this hinders the expressiveness of message passing: 
a process willing to send a name $n$ outside the module where the
restriction on $n$ is hosted is stuck.  Consequently, for two
distant agents to share a common name, this name should be allocated
at a place that is visible for both, i.e., above them in the hierarchy
of modules.
In other words, extrusion is not transparent to the user, and has to
be programmed when necessary. Of course, there are situations where
one would like to allocate a new name outside the current module. It
turns out that a corresponding primitive for remote allocation, $\new
n@m$, can be added at small cost to our implementation
(Sect.~\ref{sec:implem}).

Experiments with examples written in \kepi{} show that the idioms we
would like to be able to program are compatible with the discipline we
enforce in our formalism.  Further investigations need to be made, in
particular with larger examples, in order to understand the
possibilities offered by programming in \kepi.


\subsubsection{Modularity vs. physical deployment}

\begin{figure}[t]
  \begin{center}
    \includegraphics[width=.6\linewidth]{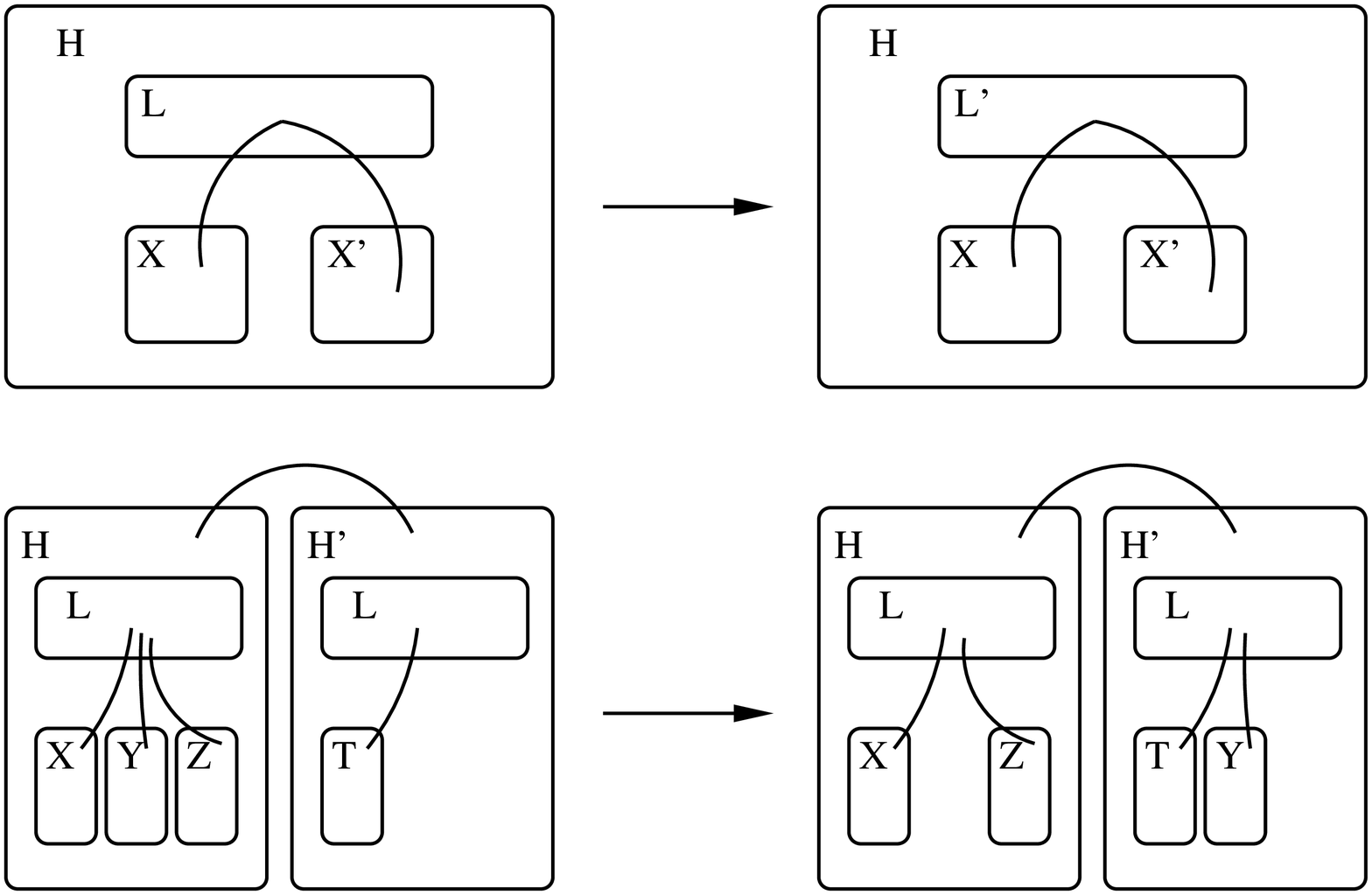}
    \caption{Some \kepi{} configurations}
    \label{fig:examples}
  \end{center}
\end{figure}

It should be noticed that the hierarchy of modules described by a
\kepi{} process does not necessarily correspond to a given mapping
from modules to physical sites. We show on Fig.~\ref{fig:examples}
some examples illustrating how we intend to use \kepi{} processes to
express typical situations in distributed programming. In the top left
diagram, a host module $H$ contains a library module $L$, as well as
two other sub-modules $X$ and $X'$. $X$ and $X'$ are connected with
module $L$ via some names residing at $H$.  Physically, this can well
correspond to a situation where $H$ and $L$ are run on the same site
(the site that provides the library), while $X$ and $X'$ are on
(possibly distinct) distant machines, and rely on remote communication
to interact with $L$. On the top right diagram, a process in $H$ has
passivated $L$ ($H$ and $L$ are co-located, physically) to replace it
with an updated version of the library ($L'$).

The bottom left diagram depicts a different scenario, based on the
same ``modularity pattern'': here, several sites provide the services
of library $L$, and welcome the users of $L$. In this configuration,
all sub-modules of a given module are executed on the same machine. It
might be the case, for load balancing purposes, that one host
passivates a local client to send it over the network to another
location: this is illustrated on the bottom right diagram.

\section{A Distributed Implementation}
\label{sec:implem}

In this section, we describe a distributed abstract machine that
implements \kepi. This machine abstracts from issues such as data
representation, to focus on the implementation of distributed
communication in the presence of passivation.  The design of this
machine has been tested on a prototypical distributed implementation,
so as to make sure that our implementation choices are reasonable
(see~\cite{kp:url}).

\subsection{Implementation Choices}

Before moving to the presentation of the formal definition of our
abstract machine, we give a high-level view of how it works. This
should help in following the more technical explanations we give in
Sec.~\ref{sec:kpam:def}.

\paragraph{Computation units.}
The first (standard) feature of our machine is that it flattens the
hierarchy of computation units: for example, each of the seven modules
composing the program of Fig.~\ref{fig:uml} is executed in its own
asynchronous \emph{location} by the machine. In order to retain the
tree structure, each location stores and maintains the list of its
children locations. As expected, when a module creates a sub-module,
the latter is spawned in a fresh location.
In order to make sure that locations can be implemented in an
asynchronous way, we let them interact only by means of (asynchronous)
messages.

\paragraph{Communications.}
The protocol for distant communication is rather standard: a process
willing to send a message sends it to the location holding the
\emph{queue} that implements the channel; accordingly, a process
willing to receive a message sends its location to the queue location
and suspends execution, so that it can be awaken when a message is
available. Using our interpretation of modules, the natural location
to run the channel queue is that of the module that created the name;
this is made possible by our choice to prevent the channel name from
being extruded out of this module: if the module gets passivated, all
of its sub-modules too, so that any process trying to communicate on
the channel will get passivated as well.



\paragraph{Passivation.}
Passivation cannot be atomic, because the hierarchy of modules has
been flattened, as described above
(this departs importantly from the machine in~\cite{BidingerSS05}).
Thus, we implement it in an incremental fashion, from the passivated
module down to its sub-modules, transitively.  Along the propagation
of a passivation session, we must handle two main sources of
interferences.
\begin{itemize}
\item First, in the case where some sub-module has already started a
  passivation, our machine gives priority to the inner passivation
  session -- the other option would have been to cancel the latter and
  let the dominating passivation proceed instead.
\item More importantly, we need to clean up running communication
  sessions in the passivated sub-modules. As explained above, this is
  not problematic for communications on names belonging to the
  passivated module. For communications on names that reside
  \emph{above} the passivated module, we use simple interactions with
  the modules owning the involved channels: status messages are used
  to query whether commitment to a communication already occurred, so
  that the computation can be either completed or aborted.
\end{itemize}


\paragraph{Distributed Implementation.} We have written an OCaml
implementation of this abstract machine~\cite{kp:url}. This
implementation exploits two libraries: one for high-level
communications, where message passing is executed either as memory
write-ups or as socket communication, depending on whether it is local
or distant; and another one for communication, thunkification and
spawning of OCaml threads, together with their sets of defined names
(to optimise the process of passivation, the data structure
implementing a module comes with a table collecting all names known by
the module). Finally, each syntactic construction of the language is
compiled into a simple function that uses the previous libraries along
the lines of the formal specification of the machine we describe
below. In particular, we do not need to manipulate explicit abstract
syntax trees at runtime.

\subsection{Formal Definition of the Abstract Machine}
\label{sec:kpam:def}

The abstract machine for the distributed execution of \kepi{}
processes is defined as a process calculus, where details about the
data structures we use in the implementation and about the protocols
at work are made apparent.

\paragraph{Configurations of the Abstract Machine.}

\begin{figure}[t]
  \centering
  \begin{align*}
    \tag{Names} &a,n,x,u,h,g,s\in Names\\
    \tag{Process Variables}  &X\in PVars \\[.4em]
    \hline\\[-.9em]
    \tag{Binder/Argument} B &::= (x) \OR (X) \OR [X] \OR \msg a \OR \msg P \OR \msg X \\
    \tag{\textbf{Process}} P &::= \nil \OR P\para P \OR (\new a) P \OR n[X] \OR \trig{a B}{P} \\[.4em]
    \hline\\[-.9em]
    \tag{Value} V &::= u^g \OR K \OR \bullet \OR \neg \\
    \tag{Request} G &::= \nil \OR G \para G \OR s(h,u,V) \OR s(\status) \\
    \tag{Local state} I &::= P \OR I \para I \OR \W^s(\trig{a B}{P},M) \OR \A^s(V) \OR \K^h_i(I) \\
    \tag{Table} \rho &::= \emptytable \OR \rho, a \mapsto u^g \OR \rho, X \mapsto K \\
    \tag{Thunk} K &::= I \dpar \rho\dpar g \\
    \tag{Messages} M &::= \nil \OR \mess{h}{I} \OR \mess{g}{G} \\
    \tag{\textbf{State}} S &::= M \OR S\para S \OR (\new a)S \OR h[K] \OR g[G] \\[.4em]
    \hline\\[-.9em]
    \tag{Contexts} \HCont^\rho_h &::= h[ [] \para I \dpar \rho\dpar g] \\
    \GCont_g &::= g[ [] \para G]
  \end{align*}
  \caption{Syntax of the abstract machine.}
  \label{fig:kpam:syntax}
\end{figure}
Fig.~\ref{fig:kpam:syntax} presents the grammar for the abstract
machine. As previously, we rely on two infinite sets of names
$(Names)$ and process variables $(PVar)$; names will be used for
\kepi{} processes as well as for various internal identifiers required
by the machine: allocated name identifiers $(u)$, logical locations
$(g,h)$, and session identifiers $(s)$. Moreover, we let $i$ range
over relative numbers $(\mathbb Z)$. Although they are presented in a
slightly different way, processes $(P)$ of the machine are essentially
the same as previously: the only difference is that we allow modules
to contain process variables only (the entry $n[P]$ disappeared, only
$n[X]$ is left). To obey this constraint, we recursively define a
pre-compilation step $\Lbag . \Rbag$ mapping processes of the calculus
(as defined in Fig.~\ref{fig:kp:syntax}) to those of the machine
(Fig.~\ref{fig:kpam:syntax}); the only non-trivial case is:
\begin{align*}
  \Lbag n[P] \Rbag &= (\new a)(\trig{a\msg{\Lbag P \Rbag}}0
  \para \rcvP{a}{X}{n[X]})
\end{align*}
\noindent (where $a$ is a fresh name).
The final translation from closed source processes to states is the
following, where $h$ and $g$ are two arbitrary distinct names:
\begin{align*}
  \llbracket P\rrbracket &= (\new h, g)(h[\Lbag
  P\Rbag\dpar\emptytable\dpar g]\para g[\nil])\enspace.
\end{align*}

At runtime, this state will evolve to a state where each module (node)
of the tree structure of $P$ is represented by a configuration of the
form $h[K]\para g[G]$, where $K=I\dpar\rho\dpar g$ is a \emph{thunk},
and:
\begin{itemize}
\item $h$ is a low-level \emph{logical} location; several such
  locations can be executed on the same \emph{physical} location;
\item $I$ is the local process being executed, together with some
  state information;
\item $g$ is the address of another associated low-level logical
  location, where names allocated by the module will be handled (with
  the help of $G$) -- we call $g$ the \emph{top-level communication
    channel} of $K$; this channel is recorded in $K$ in order to
  allocate new names.
\item $\rho$ is a \emph{table} (or environment) binding process
  variables to thunks, and names to pairs of the form $u^{g'}$: $u$ is
  a unique identifier and ${g'}$ is the channel where the name is
  handled.
\end{itemize}
In the OCaml implementation $h$ and $g$ correspond to two channels
(implemented either with shared memory or with sockets, depending on
the physical locations of the endpoints). In turn, $K$ and $G$
correspond to two threads, listening to those channels. Messages
transmitted on these channels are of the forms $I$ and $G$,
respectively.

We explain the various syntactic entries together with the operational
semantics, given below. Tables are seen as partial functions from
names and process variables to values; these functions are recursively
extended to binders and arguments (entry $B$ in the grammar) as
follows:
\begin{align*}
  \rho(\msg a) &= \rho(a) \\
  \rho(\msg P) &= (P\dpar\rho\dpar g) \quad\textrm{(for some $g\notin\fn\rho$)}\\
  \rho(\msg X) &= \rho(X)\\
  %
  %
  \rho((x)) = \rho((X)) & = \bullet \\
  \rho([X])~ &~ \textrm{is undefined}
\end{align*}
The following operation will be useful in order to extend tables:
\begin{align*}
  \rho+B\mapsto V =
  \begin{cases}
    \rho,x\mapsto u^g & \textrm{if }B=(x)\textrm{ and } V=u^g\,;\\
    \rho,X\mapsto K & \textrm{if }B=(X) \textrm{ or } B=[X],\textrm{ and } V=K\,;\\
    \rho & \textrm{otherwise.}
  \end{cases}
\end{align*}
We will also need the following substitution operation on thunks:
\begin{align*}
  K\{g\}=I\{g/g'\}\dpar \rho\{g/g'\}\dpar g \quad \textrm{ where
  }K=I\dpar\rho\dpar g'\enspace.
\end{align*}

\paragraph{Operational semantics.}

We have enough material to explain the operational semantics of the
abstract machine, which is presented in Fig.~\ref{fig:kpam:sos}. The
first three rules are standard; structural congruence is not defined
here, it contains no surprise: structural congruence of source \kepi{}
processes is inherited; parallel compositions and $\nil$ elements form
associative commutative monoids; and alpha-conversion and
scope-extrusion are allowed. We explain the other reduction rules
below; they make use of contexts ($\HCont^\rho_h,~\GCont_g$ -- see
Fig.~\ref{fig:kpam:syntax}) in order to focus on single logical
locations (we omit $\rho$, $h$ or $g$ when they are not relevant for
the rule).
\begin{figure}[!ht]
  \centering
  \small
  \begin{mathpar}
    \textbf{Contextual closure} \\
    \and
    \inferrule{ S\red S' }{ S_0\para S\red S_0\para S' } \and
    \inferrule{ S\red S' }{ (\new a)S\red(\new a) S' } \and
    \inferrule{ S\equiv S_0 \and S_0\red S'_0 \and S'_0\equiv S'}{ S\red S'} \\

    \textbf{Communication} \\
    \and
    \inferrule*[left=\RuleIReq]{ \rho(a) = u^{g} \and
      \rho(B)=V \and s\notin\fn{\HCont^\rho_h[\trig{a B}{P}]}} { \HCont^\rho_h[\trig{a B}{P}] \red (\new
      s)\bigl( \mess{g}{s(h,u,V)} \para
      \HCont^\rho_h[\W^s(\trig{a B}{P},g\msg{s(?)})] \bigr) } \and
    \inferrule[\RuleICompl]{ V\neq\neg\and \rho'=\rho+B \mapsto V}{
      \HCont^\rho[\W^s(\trig{a B}{P},\_) \para \A^s(V)] \red
      \HCont^{\rho'}[P] } \and
    \inferrule[\RuleIAbort]{ }{ \HCont[\W^s(\trig{a B}{P},\_) \para
      \A^s(\abort)] \red \HCont[\trig{a B}{P}]
    }\\

    \textbf{Routing} \\
    \inferrule*[left=\RuleIRoute]{ }{ \GCont_g[\nil] \para \mess{g}{G} \red
      \GCont_g[G]
    }\and
    \inferrule*[left=\RuleIRoute']{ }{ \HCont_h[\nil] \para \mess{h}{I} \red
      \HCont_h[I]
    }\and
    \inferrule*[left=\RuleIComm]{ }{ \GCont[s_1(h_1,u,V) \para s_2(h_2,u,\bullet)]
      \red \GCont[\nil] \para \mess{h_1}{\A^{s_1}(\bullet)} \para
      \mess{h_2}{\A^{s_2}(V)} } \and
    \inferrule*[left=\RuleIStat]{ }{ \GCont[s(h,u,V) \para s(\status)] \red
      \GCont[\nil] \para \mess{h}{\A^s(\abort)} } \\

    \textbf{Distribution} \\
    \inferrule*[left=\RuleIFresh] { a,u\notin\fn{h[I\para(\new
        a)P\dpar\rho\dpar g]}} { h[I\para(\new a)P \dpar \rho\dpar g]
      \red (\new u)h[I \para P\dpar \rho,a \mapsto u^g\dpar g] } \and
    \inferrule*[left=\RuleISpawn]{ \rho(X) = K
      \and h,g\notin\fn{\HCont^\rho_{h'}[n[X]]}} {
      \HCont^\rho_{h'}[n[X]] \red (\new h g)\bigl( h[K\{g\}] \para
      g[\nil] \para
      \HCont^\rho_{h'}[\W^h(\trig{n[X]}{n[X]},~h\msg{\K_0^{h'}(\nil)})]
      \bigr)
    }\\

    \textbf{Passivation} \\
    \and
    \inferrule*[left=\RuleIStartPass]{ M\neq\nil }{
      \HCont[\passivate{n}{X}{P} \para \W^{h}(\trig{n[X]}{n[X]},~M)]
      \red \HCont[\W^{h}(\passivate{n}{X}{P},\nil)] \para M } \and
    \inferrule*[left=\RuleIPassSess]{ M\neq\nil }{ \HCont[\K^h_i(I) \para \W^s(\trig{a B}P,M)]
      \red \HCont[\K^h_{i+1}(I \para \W^s(\trig{a B}P,\nil))] \para M} \and
    \inferrule[\RuleIDecr]{ }{ \HCont[\K^h_i(I) \para \A(V)] \red
      \HCont[\K^h_{i-1}(I \para \A(V))] } \and
    \inferrule[\RuleIPack]{ }{ h'[P \para
      \K^h_0(I) \dpar \rho\dpar g] \red \mess{h}{\A^{h'}(P\!\para\! I \dpar \rho\dpar g)} }
  \end{mathpar}
  \caption{Operational semantics for the abstract machine.}
  \label{fig:kpam:sos}
\end{figure}

\begin{itemize}
\item Communication.
  \begin{itemize}
  \item In rule \RuleIReq, a prefixed process is executed by sending a
    request $s(h,u,V)$ (for reception or for emission, according to
    the shape of $B$) on the channel that handles the communications
    taking place on $a$. A fresh session identifier, $s$, allows us to
    uniquely identify this part of the communication protocol.
    The process waiting to resume computation is stored locally in an
    element of the shape $\W^s(\trig{a B}P,M)$; message $M$ will be
    sent in case a passivation occurs before the communication
    actually take place -- we explain this protocol below.  Notice
    that this rule does not apply if $B$ is of the form $[X]$: in this
    case, $\rho(B)$ is undefined. This situation correspond to a
    passivation prefix, which is handled by rule \RuleIStartPass.
  
  \item Rule \RuleICompl{} describes how a successful completion
    message unleashes the continuation of a prefix action: a message
    $\A^s(V)$ meets an element $\W^s(\dots)$ with the same session
    identifier $(s)$: the continuation process $(P)$ is released, and,
    in case we are on the side of the receiver, a new binding is
    generated and added to the table of the local process $(\rho)$.
    
  \item In rule \RuleIAbort, a message of the shape $\A^s(\abort)$
    tells the frozen process to undo its commitment: the original
    prefixed process is installed again. As we shall see below, this
    may occur in case of a passivation.
\end{itemize}

\item Routing.
  \begin{itemize}
  \item Rules \RuleIRoute{} and \RuleIRoute' are used to transmit
    messages on both kinds of channels to their actual destination
    channel.
  \item Rule \RuleIComm{} describes how two requests for emission and
    reception, respectively, occurring on the same name identifier
    $u$, and originating from locations $h_1$ and $h_2$, meet:
    downwards acknowledgment messages are generated, with the
    identifying session names; the content of the message is
    transmitted to the receiver process.
  \item Rule \RuleIStat{} shows how messages for cancelling
    commitments $(\A^s(\abort))$ are generated: this happens when a
    message requesting communication gets caught up by a status
    message $s(?)$ (see rule \RuleIPassSess{} below).
  \end{itemize}
  
\item Distribution.  
  \begin{itemize}
  \item Rule \RuleIFresh{} shows how \kepi{} names are allocated: a
    new identifier $u$ is generated, and we store in the local table
    the information that communications on name $a$ are handled with
    identifier $u$ by the top-level communication channel $g$, which
    is associated to the current location $(h)$.

  \item Rule \RuleISpawn{} is the most complicated rule: it spawns a
    new location, in order to execute a sub-module of the current
    module. As explained above, this involves generation of two
    channels, one $(h)$ for executing the code of the sub-module, and
    another $(g)$ to handle communications on channels allocated by
    that sub-module ($g$ is the top-level communication channel).
    
    A sub-module waiting to be executed is of the form $n[X]$.  The
    local table $(\rho)$ tells which thunkified computation is
    associated to it $(K)$. This thunk has been previously linked to a
    top-level communication channel, that was used to handle
    communication channels ($K$ might be a process that has been
    running and allocating names for some time, before being
    passivated).  Since this channel cannot be used anymore (because
    the thunk may have moved physically, or been replicated) a new
    channel $(g)$ and a new thread $(g[\nil])$ have to be created, and
    we need to replace the old address with $g$ in all tables that can
    be found in $K$, whence the update $K\{g\}$.  Note that in the
    actual implementation, thanks to the use of tables, applying this
    substitution does involve inspecting the code at runtime: we only
    perform an operation on tables, which are runtime data structures
    to supervise the execution of the code.
    
    The tricky part of the rule is the element
    $\W^h(\trig{n[X]}{n[X]},~h\msg{\K_0^{h'}(\nil)})$ which is left in
    the parent location after the sub-module has been spawned.  This
    element allows us to recall that $h$ is a child of $h'$ in the
    original tree hierarchy; it will be used in two ways, in case the
    parent location wants to passivate this child, or gets passivated.
    These two cases respectively correspond to the rules
    (\RuleIStartPass,\RuleIPassSess) presented below.
  \end{itemize}

\item Passivation.
  \begin{itemize}
  \item Rule \RuleIStartPass{} initiates a passivation protocol: the
    current module wants to passivate a sub-module named $n$, and we
    know that there is such a sub-module thanks to the element of the
    form $\W(\trig{n[X]}{n[X]},~M)$. It suffices to release message
    $M$: according to rule~\RuleISpawn{}, $M$ is of the form
    $h\msg{\K_0^{h'}(\nil)}$, where $h$ is the location of the child,
    and we shall see in the remaining rules that upon reception of
    that message, the sub-module will passivate itself and send its
    thunkified version back to $h'$, the parent location. Moreover, we
    update the $\W$-element so that it will yield the expected result
    upon reception of the passivated child, through rule \RuleICompl.
    Notice that it is important to check that $M\neq\nil$: otherwise,
    in case $P$ is of the form $n[X]$, we could incorrectly apply the
    rule twice.
    
  \item Passivation gets propagated to the whole sub-tree by rule
    \RuleIPassSess{}: the $\K$-element acts as a buffer that records
    the current state of the passivation, and accepts any element of
    the form $\W(\_\,,M)$ by sending message $M$. This may correspond
    to two situations:
    \begin{itemize}
    \item- the $\W$-element denotes a sub-module, in which case the
      message tells the sub-module to passivate itself, recursively;
    \item- the $\W$-element corresponds to a pending request for
      communication. In this case, due to rule \RuleIReq, $M$ is of
      the form $g\msg{s(?)}$: by sending this \emph{status} message,
      we force $g$ to answer the request (if the completion message
      was just sent, nothing happens, the status message gets lost;
      otherwise, rule \RuleIStat{} applies, yielding an abortion
      message $h\msg{s(\neg)}$).
    \end{itemize}
    In both cases, we increment the counter stored in the
    $\K$-element: we need to record that we have to wait for an answer
    (an $\A$-message).

  \item Rule \RuleIDecr{} allows the $\K$-element to consume those
    answers, by buffering them, and decrementing its counter.
    
  \item Finally, rule \RuleIPack{} can be used after the whole state
    has been cleaned: only a source process, $P$, remains in the
    location; all $\W$ and $\A$-elements have been integrated to the
    $\K$-buffer, and the $\K$-counter is null. The buffer, the
    remaining source process, the current table and the address of the
    current name handler, $g$, are sent back to the parent location
    $(h)$, and the thread running at $h$ is killed (we could also send
    a message to the associated channel $(g)$ in order to kill the
    corresponding thread -- we omit this garbage collection
    optimisation for the sake of simplicity).
  \end{itemize}
\end{itemize}

\subsection{Related works:  abstract machines for
  distributed computation}

Several works on the implementation of process calculi for distributed
programming have focused on Ambient-like models
(Mobile~\cite{DBLP:conf/ifipTCS/FournetLS00},
Safe~\cite{GianniniSV06,HirschkoffPS05},
Boxed~\cite{DBLP:conf/esop/PhillipsYE04}). 
We have already mentioned~\cite{BidingerSS05}, which introduces an
abstract machine for the Kell calculus. The most important difference
between this work and the previous ones is the presence of the
passivation operator in the calculus, which is the source of delicate
questions when it comes to actual implementation. 
\kepi{} is related to the Kell calculus. The machine we have
introduced provides a more fine-grained presentation of distributed
interaction; in particular, passivation involves a complex distributed
protocol, while it is atomic in~\cite{BidingerSS05}.

We come back to these related studies in the next section, when
discussing the formal validation of our machine.

\section{Concluding Remarks}
\label{sec:future}

The process of the definition of \kepi{} and the study of its
implementation have raised several questions, that we are currently
investigating, or that we want to address in the future.

\subsection{Ongoing Work}

\paragraph{A type system to prevent illegal name extrusions.}
In order for the machine to work correctly, we need to make sure that
names are not extruded outside their defining module. A solution would
be to inspect the content of each message at runtime, and to block
illegal communications. To avoid the inefficiencies induced by this
approach, we want to rely on a type system to enforce statically this
confinement policy: a well-typed term never attempts to extrude a name
out of its scope.

The type system we are studying exploits an analysis of the hierarchy
of modules to detect ill-formed communications. An output \outm{a}{n}
is licit only if the restriction binding $n$ is above the one binding
$a$ in the structure of modules (or if $n$ is free in the process).
If, instead of $a$ or $n$, we have names bound to be received (as in,
e.g., $c(x)\tri \outm{a}{x}$), then the type information associated to
the transmitting channels gives an approximation of the module where
the names being communicated are allocated (intuitively, this
information boils down to \textit{``name $x$ is allocated above module
  named $m$ and under module named $k$''} -- both pieces of
information are necessary, because the name instanciating $x$ may then
be used either as medium or as object of communication).
The communication of process values follows the same ideas: in
\outm{a}{P}, we impose that all free names of $P$ should be allocated
above $a$. Consequently, in the type of both module names and of
channels over which processes are transmitted, we provide a spatial
bound of this kind on the free names of the process being executed
(resp.\ communicated).

In addition to the standard property of subject reduction, correctness
of our type system is expressed by showing that every typeable process
is \emph{well scoped}, which intuitively means that such a process
does not attempt to emit a name outside its scope. Since this property
is preserved by reduction, we can avoid checking for scope extrusions
at run time.
Although we need to experiment further with the expressiveness of our
type system, preliminary attempts show that the policy enforced by our
system is reasonable, in the sense that the examples we have in mind
can be typed in a rather natural way. We defer the presentation of the
type system, as well as the corresponding correctness proof,
to a future presentation of our work.

\paragraph{Correctness proof for the abstract machine.}


The abstract machine of Sect.~\ref{sec:implem} provides a rather
low-level description of how \kepi{} processes should be executed in a
distributed setting.
Proving its correctness, i.e., that the result of
the compilation exhibits the same behaviour as the original source
process, is a challenging task. 
Examples like~\cite{DBLP:conf/ifipTCS/FournetLS00,HirschkoffPS05}
illustrate this --- on the contrary, machines like the ones
in~\cite{BidingerS03,DBLP:conf/esop/PhillipsYE04}, by providing a more
high-level account of the implementation, make it possible to build
simpler correctness proofs.
The main difficulty is that proofs of this kind tend to be a really
large piece of mathematics; appropriate techniques are necessary to
render them more tractable, in order to be able to complete them.
This is the case in our setting, notably because the passivation
mechanism brings several technical subtleties.


The reductions of a
\kepi{} process and the execution of a machine state are described by
two transition systems.
We could hope to establish a \emph{bisimulation} result, providing
evidence that the compiled version essentially exhibits the same
behaviour as the source process. 
However, because passivation is not atomic in our setting (contrarily
to~\cite{BidingerSS05}), this is not possible.  
Indeed, consider the
following process: $m[\,a\msg{u}~|~n[b\msg{v}]\,] ~|~ \trig{m[X]}{Q}$.
The actual execution of the passivation of $m$ may go through a state
where the emission on $b$ is blocked while the one on $a$ is still
active; such a state has no counterpart in the original calculus.
Instead, correctness of our machine should be stated as a
\emph{coupled bisimulation} result~\cite{coupled:92}: although this
behavioural equivalence is weaker than plain bisimulation, it entails
operational equivalence (any \kepi{} reduction step can be simulated
by the machine, and any reduction step of the machine can be completed
into a step of the calculus). 
It can be noted that the correctness proof for the machine
of~\cite{DBLP:conf/esop/PhillipsYE04}, which comes in two parts
(soundness and completeness), is conceptually close to a 2-simulation
result, which in turn is obtained by dropping a clause from the
definition of coupled simulation.



\subsection{Future Extensions}

\paragraph{Optimisations of the machine.}

The definition and implementation of the abstract machine plays an
essential role in the design of \kepi, because it provides practical
insight on the main design decisions behind the formal model.
In addition to that, the implementation also suggests several
improvements or extensions, that we would like to study further. We
have already mentioned the primitive for remote name allocation $\new
n@m$, an operation that in principle can be encoded, but comes at a
very low cost as a primitive, given the current design of the machine.
Another direction worth investigating is how the general behaviour of
the machine can be specialised by taking into account information such
as, e.g., the fact that a whole module hierarchy runs on a single
machine.

\paragraph{Module Interfaces.}

As it is, the type system we have sketched above associates
rudimentary information to a module name, which is only related to the
regions accessed by processes running within this module (properties
like \textit{``this module has only access to references situated
  above module $m$''}). It would be interesting to define more
informative module interfaces, that would in particular describe some
aspects of the behaviour associated to the usage of the names hosted
by the module (as well as the interfaces associated to sub-modules,
recursively).
One could think for example of having the type of a channel $c$
describe the arguments that are expected to be sent on $c$.  To go
beyond that, the possibility of expressing expectations in terms of
resource access and consumption would be helpful. Also, finding
meaningful type disciplines to control usages of passivation is an
interesting question as well (a simple example is the possibility, for
implementation purposes, to be able to guarantee that a passivated
process will be used in an affine way -- that is, without duplication
-- by the passivating agent).
Existing type systems for $\pi$-calculus based formalisms, as well as
for object-oriented languages, should be sources of interesting ideas
to investigate these issues.

\paragraph{Handling (re)binding.}

In its current form, \kepi{} only makes it possible to implement
limited forms of dynamic modularity. When a module is passivated, it
can be moved around, duplicated, and computation can be resumed, as
long as the confinement constraints associated to the localisation of
restrictions are respected. In writing examples in \kepi, it appears
that it would be helpful, when passivating a module, to be able to
somehow disconnect it from some of the local resources it is using.
This would make it possible to send the passivated module outside the
scope of the names it was using, possibly to another site where
computation can be resumed after connecting again to another bunch of
resources.

Extending \kepi{} with mechanisms for dynamic (re)binding while
keeping the possibility to assert statically properties of modules
about their usage of resources is a difficult task. The work on
Acute~\cite{acute:05} may provide interesting inspiration for
this. From a more low-level perspective, we believe that the tables
that are  manipulated at run time in our machine are well-designed to
support such an extension.

\paragraph{Behavioural equivalences.}

Not only do we want to execute \kepi{} programs, but we also would
like to state and prove their properties. At a foundational level, we
would be interested in analysing the notion of behavioural equivalence
provided in \kepi, and in understanding the role of passivation in
this respect. The work of~{\cite{Lenglet2008Normal-bisimulations}}
goes in this direction (as well as~\cite{HildebrandtGM04}, in the
setting of the Homer calculus).

\bibliographystyle{entcs}
\bibliography{refs}

\begin{thebibliography}{10}
\expandafter\ifx\csname url\endcsname\relax
  \def\url#1{\texttt{#1}}\fi
\expandafter\ifx\csname urlprefix\endcsname\relax\def\urlprefix{URL }\fi
\newcommand{\enquote}[1]{``#1''}

\bibitem{BidingerSS05}
Bidinger, P., A.~Schmitt and J.~Stefani, \emph{An {A}bstract {M}achine for the
  {K}ell {C}alculus}, in: \emph{In Proc.\ {FMOODS} '05},  LNCS  \textbf{3535}
  (2005), pp. 31--46.

\bibitem{BidingerS03}
Bidinger, P. and J.~Stefani, \emph{The {K}ell {C}alculus: {O}perational
  {S}emantics and {T}ype {S}ystem}, in: \emph{In Proc.\ {FMOODS} '03},  LNCS
  \textbf{2884} (2003), pp. 109--123.

\bibitem{phd:fournet}
Fournet, C., \enquote{The {J}oin-{C}alculus: a {C}alculus for {D}istributed
  {M}obile {P}rogramming,} Ph.D. thesis, Ecole Polytechnique (1998).

\bibitem{JoCaml}
Fournet, C., F.~{Le~Fessant}, L.~Maranget and A.~Schmitt, \emph{{J}o{C}aml: {A}
  {L}anguage for {C}oncurrent {D}istributed and {M}obile {P}rogramming}, in:
  \emph{In Proc.\ {A}dvanced {F}unctional {P}rogramming 2002},  LNCS
  \textbf{2638} (2002), pp. 129--158.

\bibitem{DBLP:conf/ifipTCS/FournetLS00}
Fournet, C., J.-J. L{\'e}vy and A.~Schmitt, \emph{An asynchronous, distributed
  implementation of mobile ambients.}, in: \emph{In Proc.\ {IFIP TCS} '00},
  LNCS  \textbf{1872}, 2000, pp. 348--364.

\bibitem{GianniniSV06}
Giannini, P., D.~Sangiorgi and A.~Valente, \emph{Safe {A}mbients: {A}bstract
  machine and distributed implementation}, Sci. Comput. Program. \textbf{59}
  (2006), pp.~209--249.

\bibitem{hennessy:book:dpi}
Hennessy, M., \enquote{A {D}istributed {$\pi$}-calculus,} Cambridge University
  Press, 2007.

\bibitem{HildebrandtGM04}
Hildebrandt, T., J.~Godskesen and M.~Bundgaard, \emph{Bisimulation
  {C}ongruences for {H}omer --- a {C}alculus of {H}igher {O}rder {M}obile
  {E}mbedded {R}esources}, Technical Report TR-2004-52, Univ.\ of Copenhagen
  (2004).

\bibitem{kp:url}
Hirschkoff, D., T.~Hirschowitz, S.~Hym, A.~Pardon and D.~Pous, \emph{Abstract
  {M}achine for \kepi: {P}rototype {I}mplementation}, available from
  \texttt{http://perso.ens-lyon.fr/damien.pous/kp/kpam.tgz} (2008).

\bibitem{HirschkoffPS05}
Hirschkoff, D., D.~Pous and D.~Sangiorgi, \emph{A {C}orrect {A}bstract
  {M}achine for {S}afe {A}mbients}, in: \emph{In Proc.\ {COORDINATION} '05},
  LNCS  \textbf{3454} (2005), pp. 17--32.

\bibitem{Lenglet2008Normal-bisimulations}
Lenglet, S., A.~Schmitt and J.-B. Stefani, \emph{Normal bisimulations in
  process calculi with passivation}, Research Report RR-6664, INRIA (2008).

\bibitem{coupled:92}
Parrow, J. and P.~Sj{\"o}din, \emph{Multiway synchronizaton verified with
  woupled simulation}, in: \emph{In Proc.\ CONCUR}, 1992, pp. 518--533.

\bibitem{DBLP:conf/esop/PhillipsYE04}
Phillips, A.~T., N.~Yoshida and S.~Eisenbach, \emph{A distributed abstract
  machine for boxed ambient calculi}, in: \emph{Proc. of {ESOP}'04},  Lecture
  Notes in Computer Science  \textbf{2986} (2004), pp. 155--170.

\bibitem{Schmitt05}
Schmitt, A. and J.~Stefani, \emph{The {K}ell {C}alculus: {A} {F}amily of
  {H}igher-{O}rder {D}istributed {P}rocess {C}alculi}, in: \emph{In Proc.\
  Global Computing},  LNCS  \textbf{3267} (2005), pp. 146--178.

\bibitem{acute:05}
Sewell, P., J.~J. Leifer, K.~Wansbrough, F.~Z. Nardelli, M.~Allen-Williams,
  P.~Habouzit and V.~Vafeiadis, \emph{Acute: high-level programming language
  design for distributed computation}, in: \emph{In Proc.\ ICFP} (2005), pp.
  15--26.

\end{thebibliography}

\end{document}